\documentclass{aa}  
\usepackage{morefloats}
\usepackage{graphicx}
\usepackage{xcolor,color}
\usepackage{subfig}
\usepackage{txfonts}
\usepackage[colorlinks=true, linkcolor=blue, citecolor=blue, urlcolor=blue]{hyperref}
\usepackage{float}
\usepackage{makecell}   
\usepackage{chngcntr} 
\usepackage[title]{appendix}

\begin{document} 

\titlerunning{BAH - IV.  Discovery of a  Lensed Galaxy by  CGCG\:012-070}
\authorrunning{R. A. Riffel et al.}
\title{Blowing Star Formation Away in AGN Hosts (BAH) - III.}
\subtitle{Serendipitous discovery of a $z\sim2.9$ star-forming galaxy lensed by the galactic bulge of CGCG\:012-070 using JWST NIRSpec}

\author{Rogemar A. Riffel\inst{1,2}
\and Carlos R. Melo-Carneiro\inst{3} 
\and Gabriel Luan Souza-Oliveira\inst{2,1}
\and Rogério Riffel\inst{3} 
\and Cristina Furlanetto\inst{4} 
\and Nadia L. Zakamska\inst{5}
\and Santiago Arribas\inst{1}
\and Marina Bianchin\inst{6}
\and Ana L. Chies-Santos\inst{3}
\and José Henrique Costa-Souza\inst{2,1}
\and Maitê S. Z. de Mellos\inst{2} 
\and Michele Perna\inst{1}
\and Thaisa Storchi-Bergmann\inst{3}
}
\institute{Centro de Astrobiología (CAB), CSIC-INTA, Ctra. de Ajalvir km 4, Torrejón de Ardoz, E-28850, Madrid, Spain
\and Departamento de F\'isica, CCNE, Universidade Federal de Santa Maria, Av. Roraima 1000, 97105-900,  Santa Maria, RS, Brazil
\and Departamento de Astronomia, IF, Universidade Federal do Rio Grande do Sul, CP 15051, 91501-970 Porto Alegre, RS, Brazil
\and Departamento de Física, IF, Universidade Federal do Rio Grande do Sul, CP 15051, 91501-970 Porto Alegre, RS, Brazil
\and Department of Physics \& Astronomy, Johns Hopkins University, Bloomberg Ctr, 3400 N. Charles St, Baltimore, MD 21218, USA
\and Department of Physics and Astronomy, 4129 Frederick Reines Hall, University of California, Irvine, CA 92697, USA
}

   \date{Received July 9, 2025; accepted September 1, 2025}

 
  \abstract
  {We report the detection of a gravitationally lensed galaxy by the nearby spiral galaxy CGCG 012-070 ($z = 0.048$) using Integral Field Unit (IFU) observations with the Near-Infrared Spectrograph (NIRSpec) instrument on board the James Webb Space Telescope (JWST). The lensed galaxy is identified through the flux distributions of emission lines in the rest-frame optical, consistent with a source located at a redshift of $z\sim2.89$. The system is detected in [\ion{O}{III}]\:$\lambda\lambda4959,5007$, H$\beta$, and H$\alpha$ emission lines, exhibiting line ratios typical of a star-forming galaxy. The emission-line flux distributions reveal three distinct components, which are modeled using an elliptical power-law (EPL) mass profile for the lens galaxy. This model provides a good characterization of the source and reveals a disturbed star-forming morphology consistent with those of galaxies at cosmic noon.  This serendipitous discovery of a rare low-redshift strong lens highlights the critical role of IFU observations in expanding the lens census and advancing our understanding of galaxy mass profiles and evolution.
}


   \maketitle
%

\section{Introduction}

Strong gravitational lensing occurs when the light from a distant background galaxy is distorted and magnified by the gravitational field of a foreground  lens galaxy or galaxy cluster, aligned along the line of sight. This effect may produce multiple images or arcs of the background galaxy, with Einstein rings forming in cases of perfect alignment \citep[e.g.][]{Walsh79,Saha24}.  Strong gravitational lenses can provide valuable insights into the distribution of dark matter in galaxies and clusters \citep[e.g.][]{Tyson98,Bernabe12}, detection and characterization of high-redshift objects \citep[e.g.][]{Dye18,Chiriv20,Yue23,Abdurrouf24}, constrain cosmological parameters \citep[e.g.][]{Refsdal64,Shajib20}, and estimate the mass of supermassive black holes \citep[e.g.][]{Nightingale23,Melo-Carneiro25}, among others. 

The detection of strong gravitational lenses often involves several technical challenges, such as managing large imaging datasets and designing tools for precise identification \citep[e.g.][]{Metcalf19}. The number of identified gravitational lenses has been increasing rapidly due to the recent launch of the Euclid space telescope \citep{Euclid24} and is expected to continue to grow, thanks to surveys such as the Legacy Survey of Space and Time \citep[LSST;][]{Ivezic19}. Gravitational lenses are found mainly in early-type galaxies, with only 10–20\% in spirals \citep{Auger09}, due to their lower mass and less concentrated mass distribution. Strong lenses produced by low-redshift spiral galaxies ($z \lesssim 0.1$) are even more uncommon \citep{Treu11,ORiordan25}. Gravitational lenses in low-redshift galaxies are particularly insightful as the lensed arcs are observed in the central region of the galaxy, where the lensing primarily traces stellar mass. This, in turn, can be used as an independent constraint on the Initial Mass Function (IMF) mismatch parameter \citep[e.g.][]{Newman2017}.

Here, we report the serendipitous detection of a strongly lensed galaxy by the spiral galaxy CGCG\:012-070 ($z = 0.048$), using observations taken with the Near-Infrared Spectrograph (NIRSpec) of the James Webb Space Telescope (JWST) in integral field unit (IFU) spectroscopy mode. This paper is structured as follows. Section~\ref{sec:obs} describes the observations and data reduction. Section~\ref{sec:res} presents the main results, followed by a discussion in Section~\ref{sec:disc}. Finally, the conclusions and their implications are summarized in Section~\ref{sec:conc}.

   \begin{figure*}
   \centering
   \includegraphics[width=0.99\textwidth]{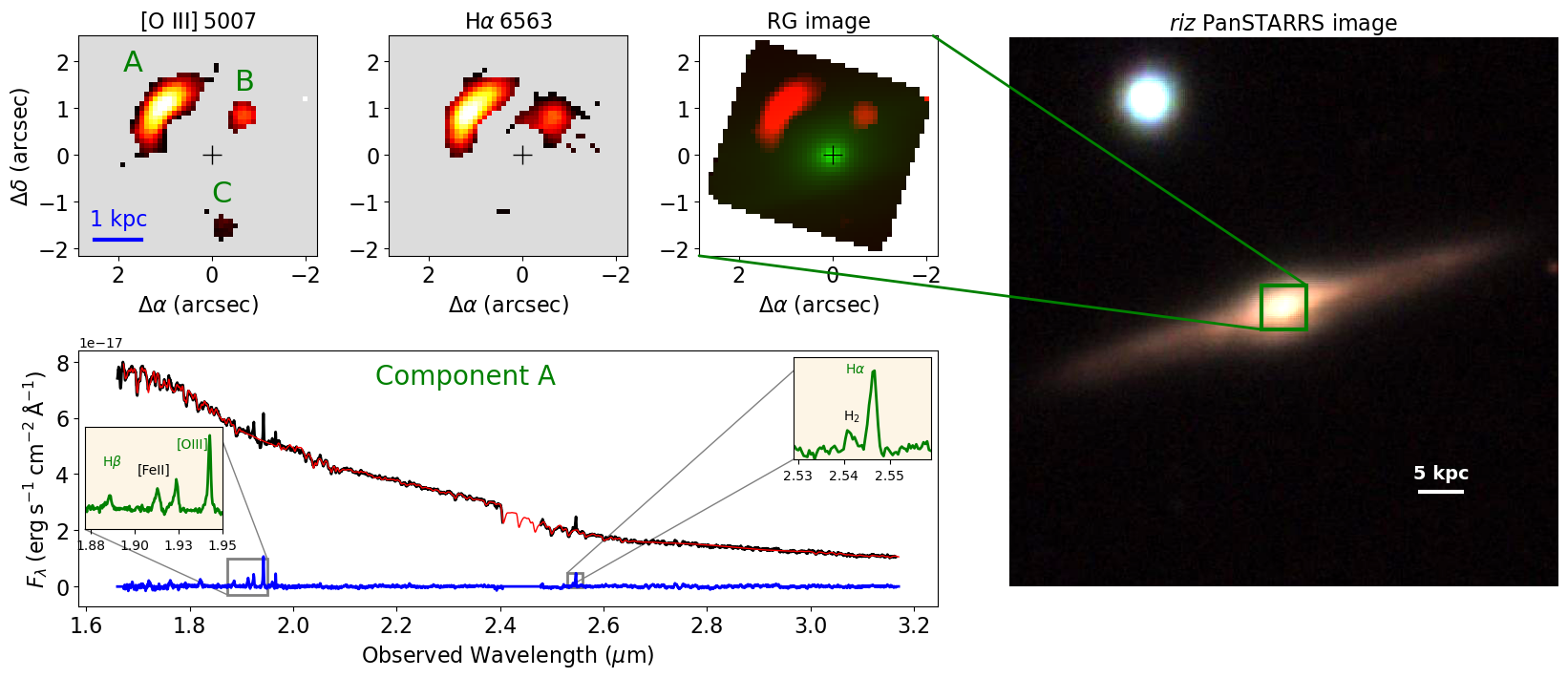} 
      \caption{The right panel shows a composite image of CGCG~012-070 obtained from the Pan-STARRS archive \citep{chambers16}, covering a region of $60 \times 60$~arcsec$^{2}$. The image combines the $r$, $i$, and $z$ bands. The green square indicates the NIRSpec field of view, which encompasses the inner region of the galaxy’s bulge.  The  small top panels show the [\ion{O}{III}] $\lambda$5007 (left) and H$\alpha$ (center) flux maps for the $z\sim2.89$ lensed object, and a composite image (right), with the image of the lensed object shown in red and the 2.0 $\mu$m continuum image of the galaxy CGCG 012-070 in green. The bottom panel shows the spectrum at the position of the lensed component A, identified in the top-left panel. The observed spectrum is shown in black, the stellar population contribution of CGCG 012-070 in red, and the gas component in blue. The insets provide a zoom-in on the [\ion{O}{III}] and H$\alpha$ regions for the lensed object. The [\ion{Fe}{II}] and H$_2$ labels, identified in black, represent Fe and H$_2$ lines from the foreground galaxy. 
              }
         \label{fig:lens}
   \end{figure*}

\section{Observations and data reduction}\label{sec:obs}

CGCG\:012-070 is a nearly edge-on ($i\approx70^\circ$) Sb galaxy and hosts a Seyfert 2 Active Galactic Nucleus \citep[AGN;][]{Kautsch2006,veron06}. It was observed as part of the Blowing Star Formation Away in AGN Hosts  \citep[BAH;][]{BAHI,BAHII}  project (PID: 1928, PI: Riffel, R. A.)
 using the JWST/NIRSpec instrument in the IFU mode \citep{Boker22}. The observations were performed using the G235H filter in combination with the F170LP grating. We processed the data using the JWST Science Calibration Pipeline \citep{bushouse_2024}, version 1.12.5, and the reference file \texttt{jwst\_1183.pmap}, following the standard pipeline stages. See \citet{deMellos25} for details.

\section{Results}\label{sec:res}

\begin{figure*}
    \centering
    \subfloat{
    \includegraphics[width=0.45\columnwidth]{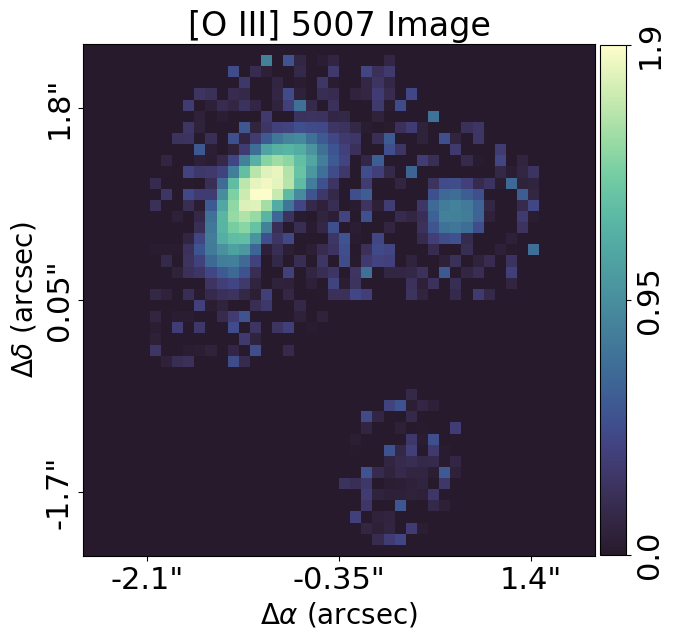}
    }
    \quad
    \subfloat{
    \includegraphics[width=0.45\columnwidth]{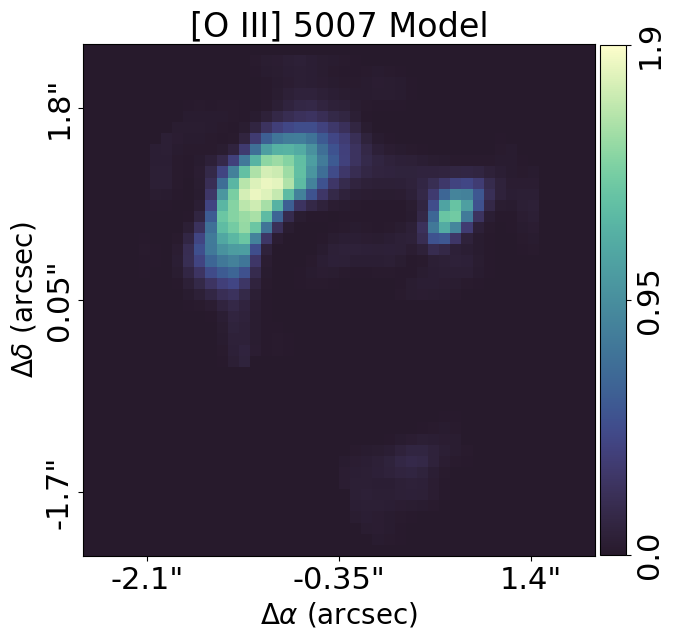}
    }
    \subfloat{
    \includegraphics[width=0.45\columnwidth,]{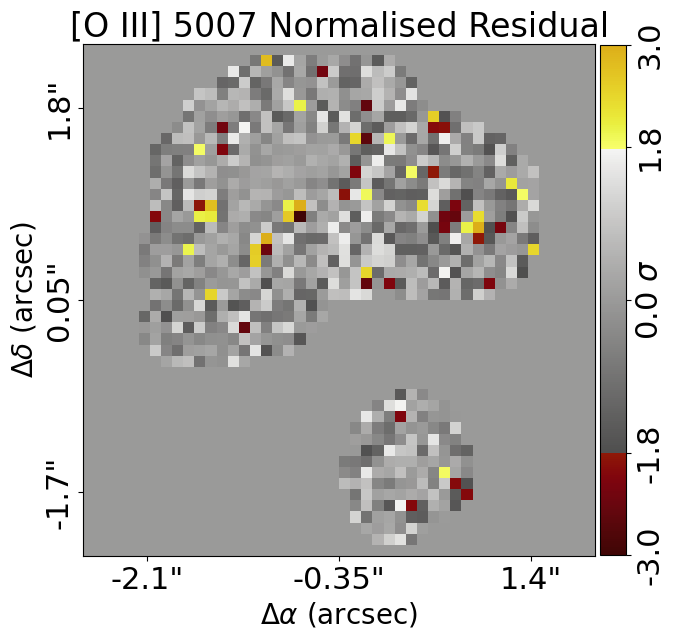}
    }
    \quad
    \subfloat{
    \includegraphics[width=0.45\columnwidth]{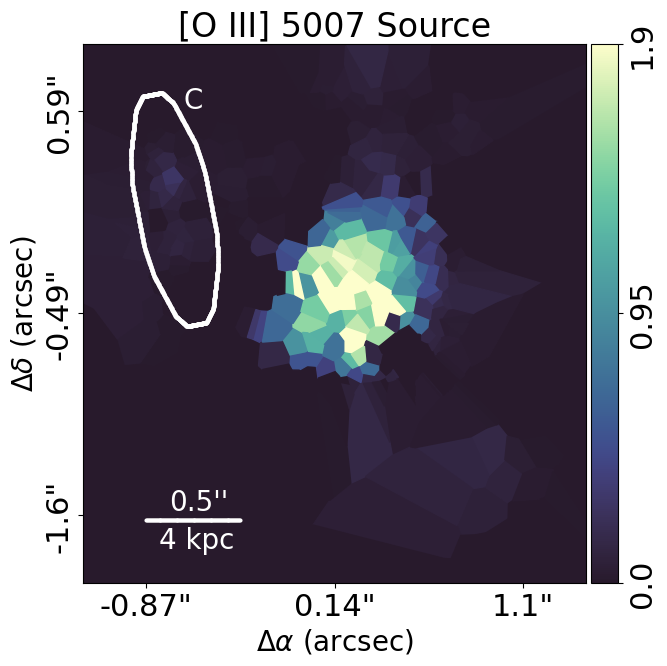}
    }
    
    \subfloat{
    \includegraphics[width=0.45\columnwidth]{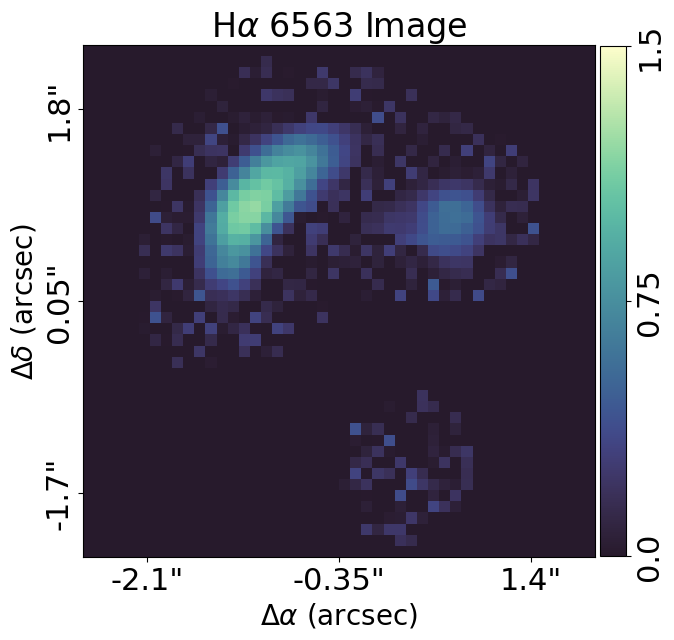}
    }
    \quad
    \subfloat{
    \includegraphics[width=0.45\columnwidth]{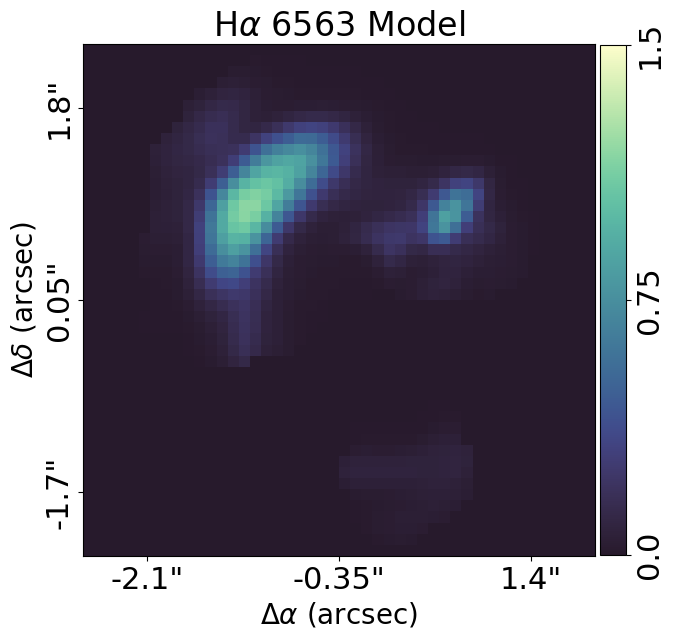}
    }
    \subfloat{
    \includegraphics[width=0.45\columnwidth,]{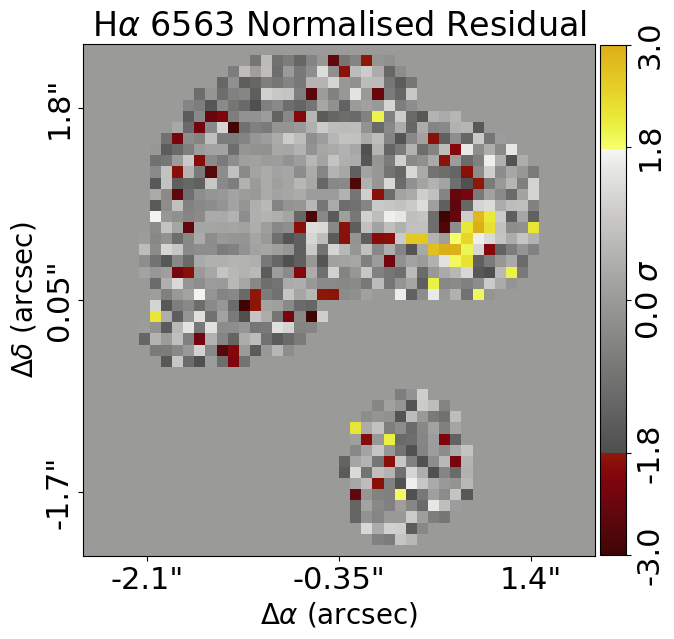}
    }
    \quad
    \subfloat{
    \includegraphics[width=0.45\columnwidth]{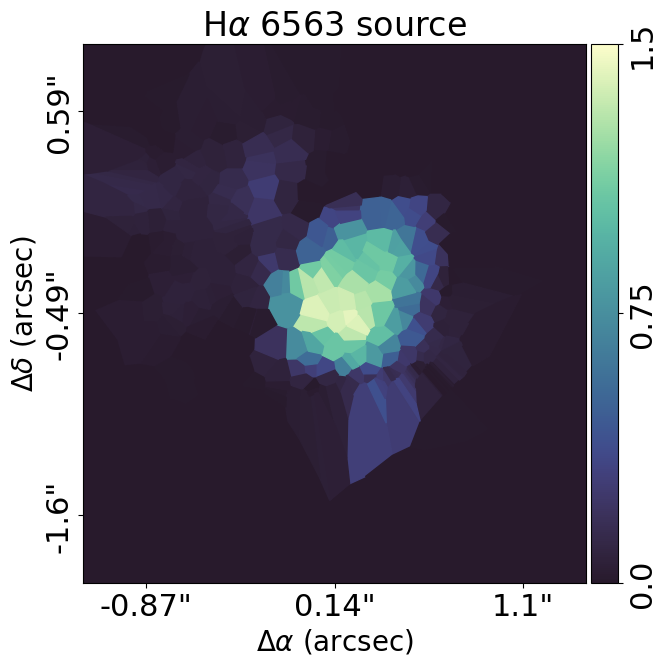}
    }    
   \caption{Highest-likelihood EPL lens model. Each row illustrates a different emission line. From left to right, we present the modified (noise-added) observed image, the model image, the normalised residuals, and the reconstructed source. The top row details the [\ion{O}{III}]\:$\lambda5007$ results, whereas the bottom row displays the H$\alpha$ reconstruction, utilising the mass model obtained from fitting the [\ion{O}{III}]\:$\lambda5007$ emission.  Fluxes are in units of $10^{-18}$ erg\,s\,cm$^{-2}$. }
    \label{fig:fidual_lens_model}
\end{figure*}

Through visual inspection of the NIRSpec data cube, we identified three components that are not associated with the galaxy but rather with a background object. These components are detected through the  H$\beta$, [\ion{O} {III}]\:$\lambda\lambda4959,5007$, and H$\alpha$ emission lines. The top panels of Fig.~\ref{fig:lens} display the observed flux distributions for [\ion{O}{III}]\:$\lambda5007$ and H$\alpha$, along with a composite image combining the [\ion{O}{III}]\:$\lambda5007$ emission (in red) and the continuum image of CGCG 012-070 (in green). The emission line maps were obtained by directly integrating the line profiles over velocity windows of 1500 km\:s$^{-1}$, after subtracting stellar population contribution obtained using the pPXF code \citep{Cappellari2023} in combination with the E-MILES templates \citep{Vazdekis2016}. 
Only spaxels where lines are detected with signal-to-noise  $snr>5$ are shown and 
 components are labeled A, B, and C in the [\ion{O}{III}]\:$\lambda5007$ flux map. Component A exhibits an arc-shaped emission, whereas components B and C appear more point-like.  The bottom panel of Fig.~\ref{fig:lens} shows the integrated spectrum for for component A. The insets highlight the regions around the emission lines of the background object. The emission lines [\ion{Fe}{II}]\:1.8298\:$\mu$m and H$_2$\:2.4237\:$\mu$m, from the foreground galaxy, are also labeled. These emission lines do not affect the flux measurements of the background object's lines, as they are outside the integration spectral windows. The spectra for the other two components are shown in Fig.~\ref{fig:spec}.

The emission components of the background source closely resemble those of strong gravitational lenses observed in other galaxies \citep[e.g.][]{Treu11,Galbany18,EuclidA,EuclidB}, supporting their interpretation as such. Using the spectra of these components, we estimated the redshift for each one based on the emission lines. Using the [\ion{O}{III}]\:$\lambda5007$ line, we derive redshifts of 2.8871, 2.8869, and 2.8844 for components A, B, and C, respectively. Similarly, from the H$\alpha$ line, we estimate redshifts of 2.8872 and 2.8873 for components A and B. 
The mean redshift is $\langle z \rangle = 2.8866 \pm 0.0011$, where the uncertainty represents the standard deviation of the mean.

\section{Discussion} \label{sec:disc}

Because of its prominence, we modelled the [\ion{O}{III}]\:$\lambda5007$ lensed emission using an elliptical power-law (EPL) mass profile for the lens galaxy, as implemented in \texttt{PyAutoLens} \citep[][]{Nightingale2018,PyAutoLens}, and with the inclusion of an external shear. Before the modelling, we added background noise sampled from a normal distribution with a zero mean and a dispersion of $\sigma_{15}= 15\%$ of the peak arc brightness, in regions where there is no [\ion{O}{III}]\:$\lambda5007$ detection \citep{Dutton2011,Barnabe2012}. Such noise is intended to suppress models that predict significant lensed features elsewhere. We assumed the uncertainty in this region to be equal to $\sigma_{15}$, and we estimated the uncertainty in the [\ion{O}{III}]\:$\lambda5007$  by assuming a Gaussian noise distribution. 
More details about the modeling strategy and systematic tests are presented in Appendix\,\ref{ap:modelling_details}.

Fig.~\ref{fig:fidual_lens_model} displays the highest-likelihood lens model, the normalised residuals, and the reconstructed [\ion{O}{III}]\:$\lambda5007$ source in the upper panels. The model accurately fits the data, particularly component A, which exhibits a higher $snr$.  Despite the overall success, significant, structured residuals are evident for component B and at the outer edges of component A. We hypothesize that this may be caused by the high elongation of the projected mass profile of the lensing disk galaxy, which a single EPL profile cannot fully capture. An EPL model is known to be insufficient in reproducing the `disky' isodensity contours (i.e., higher-order multipole moments) of a stellar disk.  The impact of this model mismatch is most pronounced for component B, as its location within the tangential critical curve makes its morphology more sensitive to complexities in the mass distribution that our model may lack. To develop more physically-motivated models, such as a multi-component mass decomposition \citep[e.g.,][]{Barnabe2012}, higher-quality imaging data would be necessary.  We then used the same mask and mass model from this fit to reconstruct the H$\alpha$ emission\footnote{For this purpose, we kept the mass model fixed and sampled only the source-plane pixelization and regularization parameters.}, which we show in the bottom panel of Fig.~\ref{fig:fidual_lens_model}. We found an effective Einstein radius for the EPL model, defined as the radius of a circle with the same area enclosed by the tangential critical curve, of $R_\text{Ein} = 1.75^{+0.03}_{-0.03}$\,arcsec ($\sim$1.8 kpc), which corresponds to a total mass within it of $M_\text{Ein}(M_\sun/10^{10}) = 7.82^{+0.25}_{-0.31}$. This value is consistent with the one we estimate from stellar population synthesis for the lens galaxy, of $M_\star(M_\sun/10^{10}) = 6.88\pm0.89$, 
 following \citet{Rogerio24}. This results in an estimated dark matter fraction of $\sim12\%$ within $R_\text{Ein}$, consistent with other low-$z$ lens galaxies \citep[e.g.][]{Newman2017,Collett2018}. We also estimate the lensing magnification factor as $\sim 1.18$.

The reconstructed [\ion{O}{III}]\:$\lambda5007$ image has a disturbed morphology. Components A and B are traced to the same spatial location in the source plane, which indicates they correspond to an image and its counter-image. This interpretation is supported by the [\ion{O}{III}]\:$\lambda5007$ velocity measurements, which show similar values for these components.  
Component C is traced back to a different location in the source plane (white ellipse) and shows a [\ion{O}{III}]\:$\lambda5007$ velocity $\sim250$ km\,s$^{-1}$ blueshifted relative to components A and B.
 Additionally, it is important to note that component C is unresolved and is detected only marginally.
 Despite tracing distinct galaxy components, such as gas rather than stars, this type of disturbed morphology is commonly observed in other lensed cosmic noon galaxies \citep{Jones10,Dye2015}.

Using the fluxes of the emission lines measured in an integrated spectrum, summing the components A and B, we obtain log\:[\ion{O}{III}]$\lambda$5007/H$\beta = 0.71\pm0.08$. Although the [\ion{N}{II}]6583 emission line is not detected, we can set an upper limit of log\:[\ion{N}{II}]$\lambda$6583/H$\alpha < -1.1$. These values are consistent with emission originating from photoionization by young stars \citep{Kewley01,Kauffmann03}, indicating that the background object is a star-forming galaxy. This is also consistent with the disturbed morphology of the gas distribution in the reconstructed source, as discussed above. 
 Using the H$\alpha$ luminosity, corrected for lensing magnification, and the relation from \citet{Kennicutt98}, we estimate a star formation rate of $\mathrm{SFR} \sim 70\,M_\odot\,\mathrm{yr}^{-1}$, which is consistent with measurements reported for galaxies at similar redshifts \citep[e.g.,][]{Forster09}.

Discovering strong galaxy-scale gravitational lenses remains challenging in ground-based surveys due to limitations imposed by atmospheric seeing. As a result, only a relatively small number of lenses were identified in the Sloan Digital Sky Survey \citep{Bolton06, Bolton08}. In contrast, space-based missions like \textit{Euclid} are now uncovering vast numbers of lenses \citep{EuclidA,EuclidB}, and the upcoming \textit{Nancy Grace Roman Space Telescope} is expected to identify even more \citep{Weiner20,Wedig25}. These efforts make it increasingly clear that the current census of single-galaxy lenses is still far from complete. Serendipitous discoveries with the JWST, particularly using its IFU capabilities, fill a unique observational niche. The spatial resolution and, crucially, the high contrast achievable within a very narrow spectral band allow the detection of lenses with Einstein radii comparable to or smaller than the bulge of the lensing galaxy. Such lenses are next to impossible to identify in broad-band imaging surveys. Such discoveries, although incidental, are crucial for expanding the sample of known lenses and for addressing key statistical questions, such as distinguishing between substructure lensing and line-of-sight halo contributions \citep[e.g.,][]{Sengul22}.

The lens reported here is produced by a low-redshift galaxy, a class of systems that is extremely rare. The probability of a galaxy at $z \sim 0.05$ producing a detectable strong lens is estimated at only 1 in 2000, so even wide-area surveys like \textit{Euclid} are expected to find just a few (4–20) such systems \citep{ORiordan25}.
 To date, only five strong lensing systems with lens galaxies at $z < 0.05$ have been identified, including the one presented here, which is the second hosted by a spiral galaxy. Its lensed source is the most distant among those associated with low-redshift lenses (Table~\ref{table:lowlz_lenses}). Discovering and building a census of strong lenses at very low redshift is crucial for anchoring models of galaxy evolution, as such systems enable direct measurements of inner mass profiles on kiloparsec scales and provide key constraints on cosmological predictions for the redshift evolution of galaxy density slopes \citep{Sahu24}.
This underscores both the rarity and the significance of discovering a low-redshift lens through serendipitous IFU observations.

\section{Conclusions} \label{sec:conc}

This work presents a remarkable discovery of a gravitational lens system with JWST, confirming the origin of its lensed [\ion{O}{III}]\:$\lambda5007$ and H$\alpha$ emissions. We estimate a redshift of $z \approx 2.89 $ background source based on the central wavelengths of the observed emission lines. Additionally, the intensity ratios of these lines indicate that the primary ionization source of the gas is star formation. We modeled the lens system using an EPL profile, which effectively identified components A and B as a confirmed image and its counter-image. Furthermore, the model revealed a disturbed, star-forming source morphology consistent with cosmic noon galaxies. While our model provides a strong initial characterization, future deep imaging and spectroscopy will enable more detailed investigations of the lens and source objects, including constraining the IMF, but also will allow us to combine lens kinematics to precisely separate stellar and dark matter contributions within the lens galaxy.

\begin{acknowledgements}
This work is based on observations made with the NASA/ESA/CSA James Webb Space Telescope. The data were obtained from the Mikulski Archive for Space Telescopes (MAST) at the Space Telescope Science Institute, which is operated by the Association of Universities for Research in Astronomy, Inc., under NASA contract NAS 5-03127 for JWST. These observations are associated with program \#1928. The complete dataset can be accessed at MAST platform, through \url{https://doi.org/10.17909/tazj-hp44}. We thank the referee for valuable suggestions. ChatGPT (GPT-4.5) was used to improve sentence wording. RAR, GLSO, JHCS and MZM acknowledge the support from the Conselho Nacional de Desenvolvimento Científico e Tecnológico (CNPq; Projects 303450/2022-3, 403398/2023-1, and 441722/2023-7) and the Coordenação de Aperfeiçoamento de Pessoal de Nível Superior (CAPES; Project 88887.894973/2023-00, and Finance code 0001). RR acknowledges support from  CNPq (Proj. 445231/2024-6,311223/2020-6, 404238/2021-1, and 310413/2025-7), FAPERGS, (Proj. 19/1750-2 and 24/2551-0001282-6) and CAPES (88881.109987/2025-01). CF acknowledges the support from CNPq Project 315421/2023-1. ACS acknowledges support from FAPERGS (grants 23/2551-0001832-2 and 24/2551-0001548-5), CNPq (grants 314301/2021-6, 312940/2025-4, 445231/2024-6, and 404233/2024-4), and CAPES (grant 88887.004427/2024-00).
\end{acknowledgements}

\bibliography{aa56329-25}

\begin{appendix}

\section{The lens galaxy: CGCG\:012-070}

   \begin{figure*}[ht]
   \centering
   \includegraphics[width=0.9\textwidth]{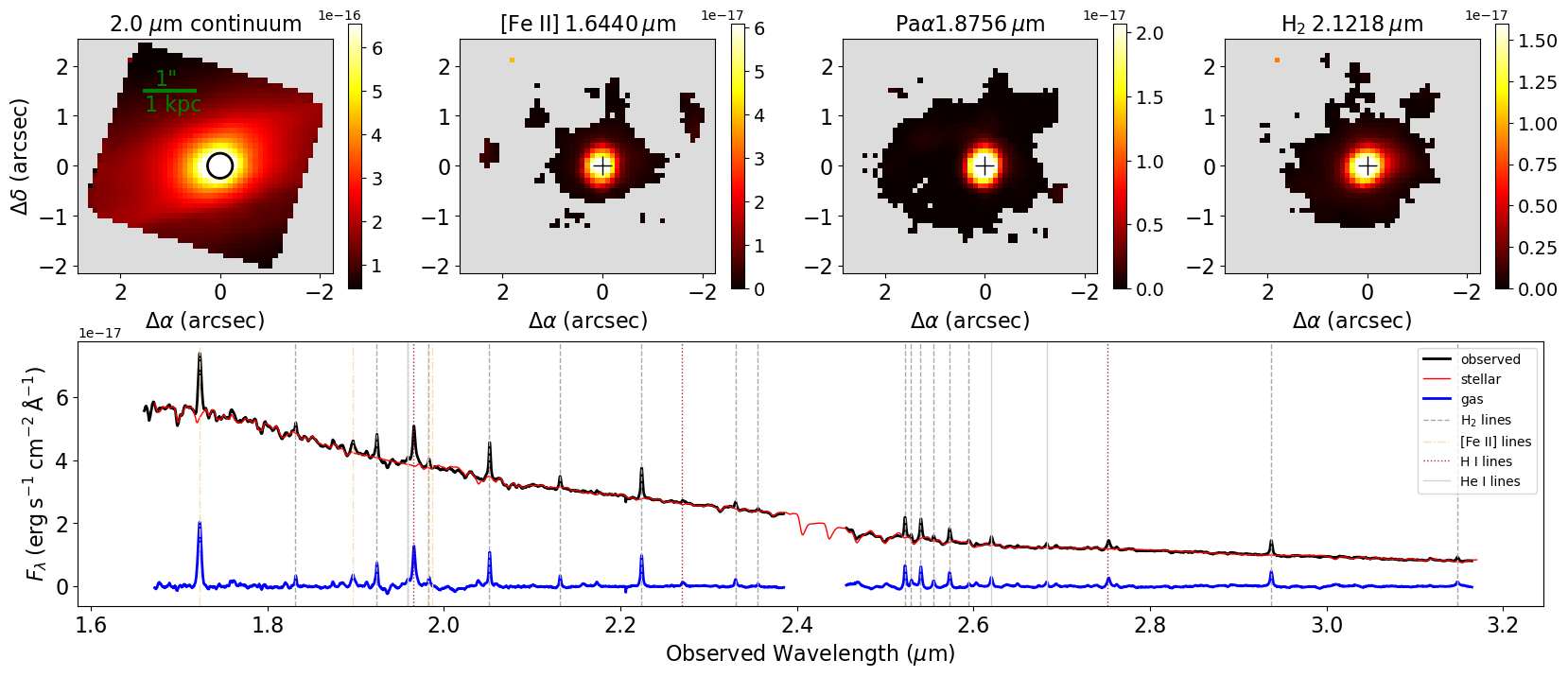}
      \caption{{\bf The CGCG 012-070 galaxy.} The top-left panel shows a 2.0$\mu$m continuum image of CGCG 012-070, obtained from the NIRSpec data cube. The other top panels display the flux distributions of the  [\ion{Fe}{II}]\:1.6440\:$\mu$m, Pa$\alpha$, and H$_2$\:2.1218$\:\mu$m emission lines. Gray regions represent masked out areas where the $snr<3$ and regions outside the NIRSpec FoV. The cross marks the position of the continuum peak. The bottom panel presents the observed spectrum (in black), integrated over a 0\farcs25 radius nuclear aperture (black circle in the top-left panel). The red curve represents the stellar population (SP) model, while the blue curve shows the gas contribution (observed - SP). Vertical lines mark the position of the main emission lines.
              }
         \label{fig:cgcg}
   \end{figure*}

In Fig.~\ref{fig:cgcg}, we present an image of the continuum at 2.0 $\mu$m extracted from the NIRSpec data cube, along with flux maps for the [\ion{Fe}{II}]\:1.6440\:$\mu$m, Pa$\alpha$, and H$_2$\:2.1218$\:\mu$m emission lines. These maps were created by integrating the spectra over a 1500 km\:s$^{-1}$ window centered on each emission line, after subtracting stellar population contribution obtained using the pPXF code \citep{Cappellari2023} in combination with the E-MILES templates \citep{Vazdekis2016}.  The bottom panel shows the integrated nuclear spectrum integrated within a 0\farcs25 radius aperture in black, while the underlying stellar continuum is shown in red and the gas emission spectrum is shown in blue.  
The continuum image follows the same orientation as the large-scale disc, elongated along position angle PA$\approx$107$^\circ$ \citep{Adelman08}.  The emission-line flux distributions peak at the galaxy's nucleus, with weaker extended emission detected up to $\sim$2$^{\prime\prime}$ from it.  The investigation of the origin of gas emission and kinematics is beyond the scope of this paper and will be addressed in a forthcoming study.

\section{Spectra of components B and C}

\begin{figure*}
   \centering
   \includegraphics[width=0.89\textwidth]{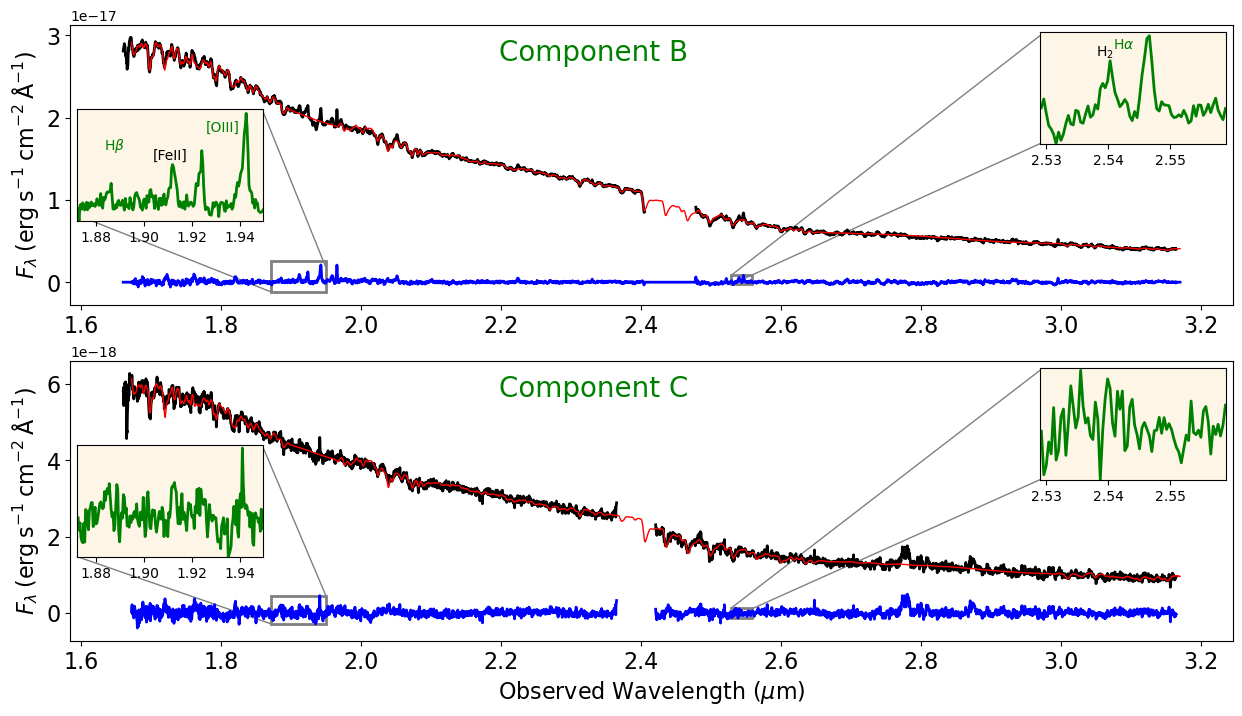} 
      \caption{Integrated spectra for components B (top) and C (bottom). In each panel, the observed spectrum is shown in black, the stellar population contribution of CGCG 012-070 in red, and the gas component in blue. The insets provide a zoom-in on the [\ion{O}{III}] and H$\alpha$ regions for the lensed object.  The [\ion{Fe}{II}] and H$_2$ labels identify emission lines from the foreground galaxy. }              
         \label{fig:spec}
   \end{figure*}

Figure~\ref{fig:spec} shows the integrated spectra (in black) of the lensed components B and C, along with the corresponding stellar population models (in red) for CGCG~012-070. The residual spectra, shown in blue, were obtained by subtracting the stellar contribution from the integrated spectra. All observed spectra were extracted by summing the spaxels where the [\ion{O}{III}] emission line is detected with $snr>5$ in each component, identified in Fig.~\ref{fig:lens}. 

\section{Some details of the lens modeling}\label{ap:modelling_details}


\begin{figure*}
    \centering
    \subfloat{
    \includegraphics[width=0.50\columnwidth]{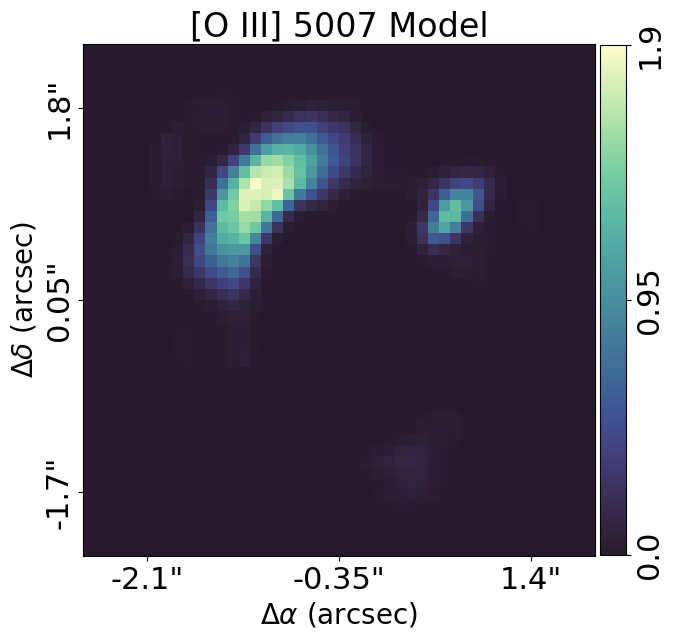}
    }
    \subfloat{
    \includegraphics[width=0.50\columnwidth,]{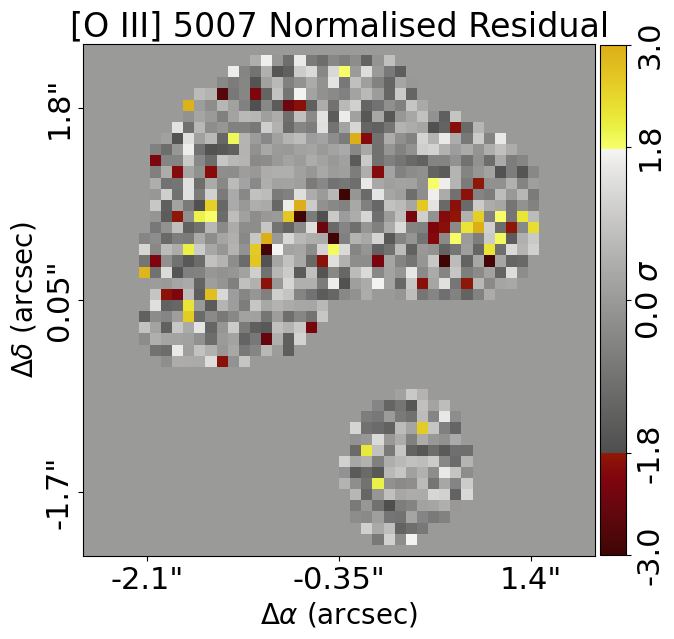}
    }
    \quad
    \subfloat{
    \includegraphics[width=0.50\columnwidth]{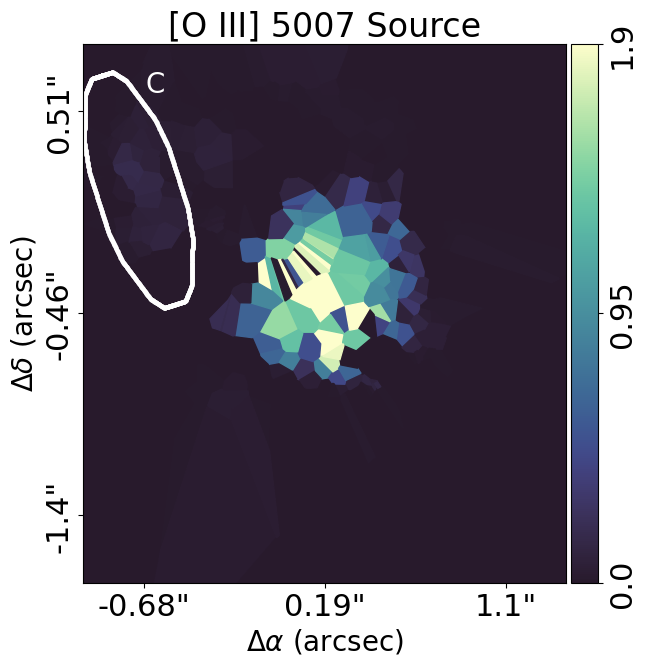}
    }
    \quad
    \subfloat{
    \includegraphics[width=0.50\columnwidth]{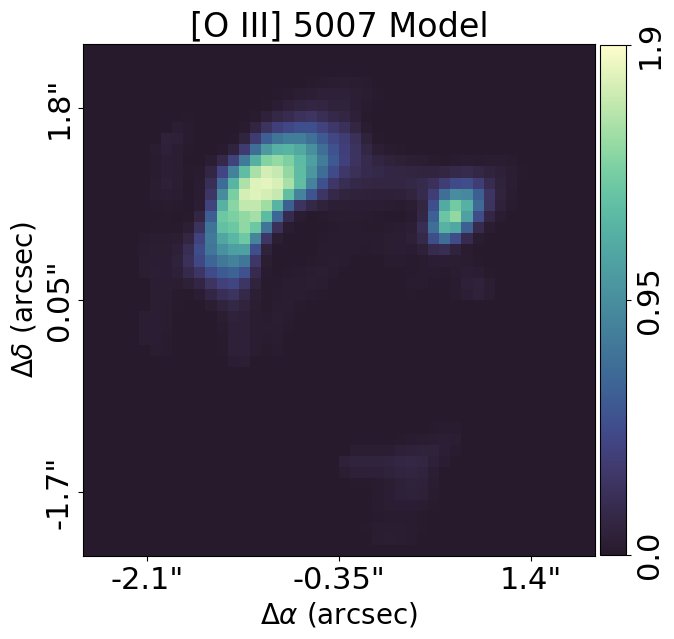}
    }
    \subfloat{
    \includegraphics[width=0.50\columnwidth,]{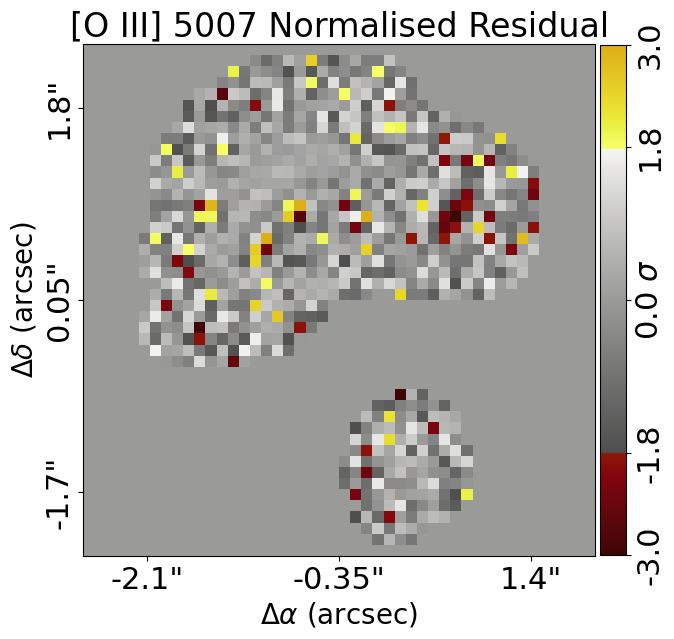}
    }
    \quad
    \subfloat{
    \includegraphics[width=0.50\columnwidth]{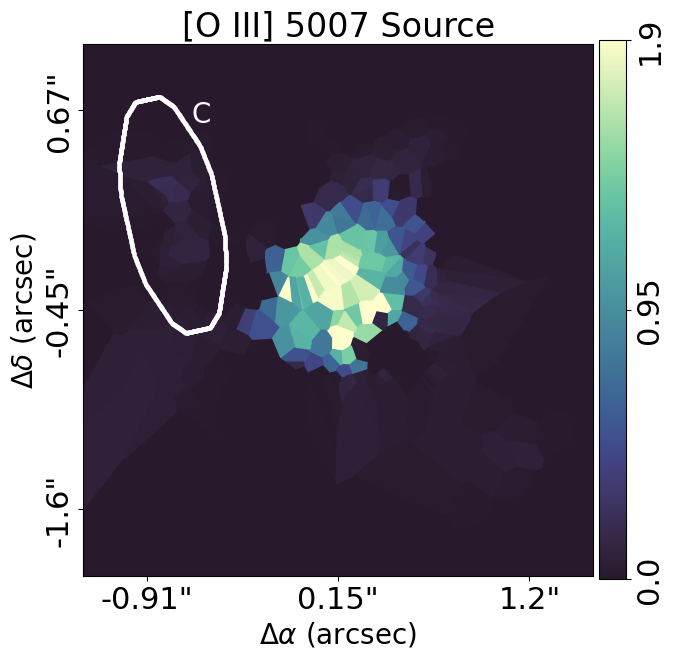}
    }    
   \caption{Highest-Likelihood EPL Lens Models for Alternative Datasets. Each row illustrates a different assumption for the noise added to empty regions. From left to right, we present the model image, normalized residuals, and reconstructed source. The top row shows the dataset where the modified image uses a $\sigma_{10}$ dispersion, while the bottom row corresponds to a $\sigma_{20}$ dispersion. Fluxes are in units of $10^{-18}$ erg\,s$^{-1}$\,cm$^{-2}$. }
    \label{fig:alternative_models}
\end{figure*}

We performed the modelling using the open source lens modelling package \texttt{PyAutoLens} \citep{Nightingale2018}. Our approach utilized the SLaM (Source, Light, and Mass) pipelines distributed with \texttt{PyAutoLens}, similar to the methods employed in other works \citep[e.g.,][]{Cao2022,Etherington2022}, but with modifications to match our data. The pipeline workflow is summarized below:

\begin{enumerate}
    \item Initialize: We began by creating a basic model for the lens mass and source light distribution using parametric profiles. The lens mass is assumed to be a singular isothermal ellipsoid (SIE)+shear, and the source's brightness was modeled by a Sersic component. 
    \item Lens Refinement and Source Pixelization: Using the results from the initial step, we re-sampled the lens mass model parameters alongside a pixelized source model. We employed a Voronoi mesh grid with Natural Neighbor interpolation \citep{Sibson1981} and brightness-adaptive regularization to smooth the reconstruction based on source luminosity. Priors on the mass parameters were updated using \texttt{PyAutoLens}'s default prior passing\footnote{Additional details are available in the online documentation for PyAutoLens: \url{https://github.com/Jammy2211/autolens_workspace}.}.
    \item Source Pixelization Refinement: We refined the source plane reconstruction using the results from the previous phase. The mass model was fixed to its highest-likelihood parameters, while the source plane parameters were re-sampled. The source was reconstructed using a brightness-adaptive Voronoi mesh grid with Natural Neighbor interpolation and brightness-adaptive regularization. In \texttt{PyAutoLens} notation, this corresponds to a \texttt{KMeans} mesh grid, with \texttt{VoronoiNN} pixelization, and \texttt{AdaptiveBrightnessSplit} regularization. During this reconstruction, we fixed the number of \texttt{KMeans} clusters to 500.
    
    \item  Lens Mass Refinement: With the source reconstruction fixed to the highest-likelihood model from the previous step, we refined the mass model. The lens mass profile was updated to an EPL + shear, allowing for greater complexity. Again, the priors are updated using the default prior passing from \texttt{PyAutoLens}. 
\end{enumerate}

\begin{table}
\centering
\caption{Inferred median and $1\sigma$ credible intervals for model parameters.  }
\renewcommand{\arraystretch}{1.8}
\label{table:posterior}
\begin{tabular}{lccc} 
\hline
Parameter  & Model $\sigma_{10}$ & \makecell{Model $\sigma_{15}$ \\ (fiducial)} & Model $\sigma_{20}$\\
\hline
$x_0[\arcsec]$                              &   $-0.05^{+0.01}_{-0.01}$ &  $-0.05^{+0.01}_{-0.01}$ &  $-0.03^{+0.01}_{-0.01}$   \\
$y_0[\arcsec]$                              &   $-0.10^{+0.01}_{-0.02}$ &  $-0.14^{+0.02}_{-0.03}$ &  $-0.19^{+0.03}_{-0.03}$   \\
$\epsilon_1$                                &   $0.75^{+0.01}_{-0.01}$  &  $0.74^{+0.02}_{-0.02}$  &  $0.68^{+0.02}_{-0.02}$   \\
$\epsilon_2$                                &   $-0.07^{+0.01}_{-0.02}$ &  $-0.10^{+0.03}_{-0.03}$ &  $-0.16^{+0.04}_{-0.04}$   \\
$\theta^{\text{lens}}_\text{Ein}[\arcsec]$  &   $3.00^{+0.11}_{-0.10}$  &  $2.97^{+0.16}_{-0.13}$  &  $2.69^{+0.10}_{-0.09}$    \\
$\gamma^{\text{lens}}$                      &   $2.19^{+0.03}_{-0.04}$  &  $2.27^{+0.03}_{-0.04}$  &  $2.37^{+0.04}_{+0.04}$    \\
$\epsilon^{\text{sh}}_1$                    &   $-0.09^{+0.02}_{-0.01}$ &  $-0.06^{+0.01}_{-0.01}$ &  $-0.07^{+0.01}_{+0.01}$   \\
$\epsilon^{\text{sh}}_2$                    &   $-0.04^{+0.03}_{-0.02}$ &  $-0.12^{+0.03}_{-0.02}$ &  $-0.18^{+0.02}_{-0.02}$   \\
$R_\text{Ein}[\arcsec]$                     &   $1.76^{+0.02}_{-0.02}$  &  $1.75^{+0.03}_{-0.03}$  &  $1.67^{+0.03}_{-0.04}$    \\ 
$M_\text{Ein}[M_\sun/10^{10}]$              &   $7.85^{+0.22}_{-0.16}$  &  $7.82^{+0.25}_{-0.31}$  &  $7.09_{-0.30}^{+0.23}$    \\[1ex]
\hline
\end{tabular}
\tablefoot{
 Parameters are, in order, lens potential centre $(x_0,y_0)$, lens mass elliptical components $(\epsilon_1,\epsilon_2)$, EPL Einstein radius $\theta^{\text{lens}}_\text{Ein}$, EPL density slope $\gamma^{\text{lens}}$, elliptical shear components $(\epsilon^{\text{sh}}_1, \epsilon^{\text{sh}}_2)$,  effective Einstein radius $R_\text{Ein}$, and the total mass within it $M_\text{Ein}$. Columns represent individual datasets or models (refer to text for details).
}
\end{table}

This pipeline concludes with the fitting of the EPL+shear model, which is the final model used for the analysis presented in this work. The convergence of the EPL mass profile is given by  \citep[][]{Tessore2015}:

\begin{equation}
    \kappa(\xi) = \frac{\left( 3 - \gamma^{\text{lens}}\right)}{1 + q^{\text{lens}}} \left(\frac{\theta^{\text{lens}}_\text{Ein}}{\xi}\right)^{\gamma^{\text{lens}} -1}.
\end{equation}
Here, $q^{\text{lens}}$ is the axis ratio (minor-to-major axis), and $\xi$ is the elliptical coordinate, defined as $\xi = \sqrt{x^2 + (y/q^{\text{lens}})^2}$. The parameter $\theta^{\text{lens}}_\text{Ein}$ is Einstein radius in units of arcsec, and $\gamma^{\text{lens}}$ is the mass density slope; this profile reduces to a SIE when $\gamma^{\text{lens}} = 2$. 

Additionally, the mass position angle, $\phi^{\text{lens}}$, measured counter-clockwise from the positive $x$-axis, can be incorporated using the elliptical components:

\begin{equation}
    \epsilon_1 = \frac{1 - q^{\text{lens}}}{1 + q^{\text{lens}}}\sin{2\phi^{\text{lens}}}, \quad \quad  \epsilon_2 = \frac{1 - q^{\text{lens}}}{1 + q^{\text{lens}}}\cos{2\phi^{\text{lens}}}.
\end{equation}
It is worth mentioning that the Einstein radius $\theta^{\text{lens}}_\text{Ein}$ in this equation differs from the {\it effective} Einstein radius as defined by \citet{Meneghetti2013}. The {\it effective} Einstein radius, which is the one applied in our analysis, corresponds to the radius of a circle having the same area as the region enclosed by the tangential critical curve.

The lensing external shear is parametrized by its elliptical components, $(\epsilon^{\text{sh}}_1, \epsilon^{\text{sh}}_2)$, as follows:

\begin{equation}\label{eq:shear}
    \gamma^{\text{sh}} = \sqrt{{\epsilon^{\text{sh}}_1}^2 + {\epsilon^{\text{sh}}_2}^2}, \quad \tan{\left(2\phi^{\text{sh}}\right)} = \frac{\epsilon^{\text{sh}}_2}{\epsilon^{\text{sh}}_1}, 
\end{equation}
where $\gamma^{\text{sh}}$ and $\phi^{\text{sh}}$ are the shear amplitude and shear angle, measured counter-clockwise from north, respectively. 

To account for PSF blurring effects in the lens modeling, we used the \texttt{STPSF}\footnote{\url{https://stpsf.readthedocs.io/en/latest/index.html}} \citep{STPSF} tool to reconstruct the PSF at each wavelength across the [\ion{O}{III}]~$\lambda$5007 and H$\alpha$ emission lines. For each line, we computed a final PSF by taking the median of the wavelength-dependent reconstructions, resulting in two stacked PSFs --- one for [\ion{O}{III}] and one for H$\alpha$ --- which were then used in the modeling. The FWHM for the [\ion{O}{III}] emission is $\sim0.16\arcsec$ ($\sim0.165$\,kpc), while for the H$\alpha$ emission it is $\sim0.17\arcsec$ ($\sim0.165$\,kpc), at the lens redshift. 

We performed the sampling using the nested sampler \texttt{dynesty} \citep{Speagle2020}. When necessary, we adopted a cosmology consistent with \cite{Planck_2015}. The median of each parameter's one-dimensional marginalized posterior distribution, along with uncertainties corresponding to the $16^\text{th}$ and $84^\text{th}$ percentiles for our models, are presented in Table~\ref{table:posterior}.

We investigated the impact of the introduced noise level on empty regions on our modeling results. While our fiducial model assumes a dispersion of $\sigma_{15}=15\%$ of the peak arc brightness for the introduced noise, we performed additional tests using dispersion values of $\sigma_{10}$ and $\sigma_{20}$. The results of these alternative datasets are presented in Fig.~\ref{fig:alternative_models}, with the median posterior distributions of the parameters detailed in Table~\ref{table:posterior}. Overall, our results demonstrate consistency across the tested noise levels, but we found that the data were best fitted when a dispersion of $\sigma_{15}$ was applied (see the normalized residual maps at the B region).

\section{Strong gravitational lens around low redshift galaxies}

In Table~\ref{table:lowlz_lenses}, we list strong gravitational lens systems associated with galaxies at $z < 0.05$ reported in the literature. Among these, the lensed source presented in this work is the most distant, and the system is only the second known case involving a spiral lens galaxy at such low redshift.  

\begin{table}
\centering
\caption{Strong lensed systems by $z<0.05$ galaxies. }
\renewcommand{\arraystretch}{1.4}
\label{table:lowlz_lenses}
\begin{tabular}{lcccc} 
\hline
Lens Galaxy & $z_\mathrm{lens}$ & HT & $r_\mathrm{Ein}$ (kpc) & $z_\mathrm{src}$\\
\hline
CGCG\,378$-$015$^{\rm a,b}$ & 0.039 & S & 0.8 & 1.695 \\
ESO\,325$-$G004$^{\rm c}$ & 0.034 & E & 1.9 & 2.141 \\
ESO\,286$-$G022$^{\rm c}$ & 0.031 & E & 1.5 & 0.926\\
NGC\,6505$^{\rm d}$ & 0.042 & E & 2.1 & 0.406  \\
CGCG\:012-070$^{\rm e}$ & 0.048 & S & 1.8 & 2.887 \\
\hline
\end{tabular}
\tablefoot{ Columns list the lens galaxy name, redshift of the lens ($z_\mathrm{lens}$), morphological classification (HT; S: spiral; E: elliptical), Einstein radius ($r_\mathrm{Ein}$), and redshift of the background source ($z_\mathrm{src}$). References: a) \citet{Huchra1985} b) \citet{Trott10}; c) \citet{Smith15}; d) \citet{ORiordan25}, e) This work.  }
\end{table}

\end{appendix}
 \end{document}